\documentclass[twocolumn,showpacs,preprintnumbers,amsmath,amssymb,pre,superscriptaddress]{revtex4}
\usepackage{color}

\usepackage{graphicx}
\usepackage{dcolumn}
\usepackage{bm}
\usepackage{subfigure}
\usepackage{amssymb}
\renewcommand{\vec}{\boldsymbol}

\begin{document}

\title{Higher harmonics of the magnetoplasmon in strongly coupled \\Coulomb and Yukawa systems}

 \author{T. Ott}
 \affiliation{%
     Christian-Albrechts-Universit\"at zu Kiel, Institut f\"ur Theoretische Physik und Astrophysik, Leibnizstra\ss{}e 15, 24098 Kiel, Germany
 }%
 
 \author{M. Bonitz}%
 \affiliation{%
     Christian-Albrechts-Universit\"at zu Kiel, Institut f\"ur Theoretische Physik und Astrophysik, Leibnizstra\ss{}e 15, 24098 Kiel, Germany
 }%
\author{P.~Hartmann}
\affiliation{Research Institute for Solid State Physics and Optics, Hungarian Academy of Sciences, P. O. Box 49, H-1525 Budapest, Hungary}
 \author{Z.~Donk\'o}
\affiliation{Research Institute for Solid State Physics and Optics, Hungarian Academy of Sciences, P. O. Box 49, H-1525 Budapest, Hungary}

\date{\today}

\begin{abstract}
The generation of higher harmonics of the magnetoplasmon frequency which 
has recently been reported in strongly coupled two-dimensional Yukawa systems is investigated in detail and, in addition, 
extended to two-dimensional Coulomb systems. 
We observe higher harmonics over a much larger frequency range than before and compare the theoretical prediction with the simulations. 
The influence of the coupling, structure, and thermal energy on the excitation of these modes is examined in detail. 
We also report on the effect of friction on the mode spectra to make predictions about the experimental observability of this new effect. 

\end{abstract}

\maketitle

\section{Introduction}

The behaviour of two-dimensional (2D) systems is of continuing interest in many fields of physics as the 
reduced number of dimensions give rise to a number of peculiar properties. 
In highly correlated systems, where the potential energy due to the interaction dominates over the thermal energy (for recent overviews see \cite{Donko2008,bonitz_rpop10}), 2D many-particle systems exhibit a strongly collective behaviour which manifests itself, e.g., in anomalous transport properties of 2D liquids~\cite{Liu2008a,Ott2008,Donk'o2009,Ott2009c,Ott2009b}. Correlational effects are also responsible for additional shear mode excitations of 2D liquids~\cite{Golden2000,Kalman2000} which do not occur in weakly coupled 
systems but have been experimentally observed in strongly coupled dusty plasmas~\cite{Nunomura2005,Piel2006}. 

Subjecting a 2D many-particle system to a perpendicular magnetic field gives rise to yet another line of research 
into the physics of low-dimensional systems~\cite{Ranganathan2002}. The magnetic field effectively ``mixes'' longitudinal and 
transverse excitations, leading to two hybrid modes, the magnetoplasmon and the magnetoshear, which have 
in recent times been studied in Coulomb and Yukawa systems~\cite{Uchida2004,Ranganathan2005,Hou2009c}. 
These modes are well understood from a theoretical perspective, including descriptions based on the quasi-localized charge 
approximation (QLCA)~\cite{golden_1992,Jiang2007} or in harmonic lattice approximations~\cite{Kalman2000,Hou2009c}. 

However, besides these established modes, 2D Yukawa systems at strong coupling have recently been found to support 
additional high-frequency modes. These modes appear as higher harmonics of the magnetoplasmon 
in non-dissipative 2D Yukawa systems~\cite{Bonitz2010} and are reminiscent of the classical Bernstein modes~\cite{Bernstein1958}. 
Unlike these, however, the observed higher harmonics are not a pure magnetic effect but are, additionally, 
fundamentally affected by 
the strong correlations between particles, as was demonstrated in Ref.~\cite{Bonitz2010}.

In this work, we put the theoretical predictions of our earlier work~\cite{Bonitz2010} to the test by using more detailed simulations 
and observing considerably more high-frequency modes. In addition, we include the Coulomb case of vanishing screening 
in our analysis, and demonstrate that the higher harmonics are also generated under these circumstances, which are not only 
quantitatively but also qualitatively different from screened-interaction systems. Finally, we assess the 
possibility of experimental verification by investigating the relative intensities of the modes and including dissipative 
effects in our simulations.

The remainder of this article is structured as follows: In Section~\ref{sec:model}, we introduce our model and 
give details of the numerical procedure and the system of units. Section~\ref{sec:collective} introduces the 
longitudinal and transverse fluctuation spectra and the means by which they are computed. The results 
for non-dissipative and dissipative systems are presented in Section~\ref{sec:results} 
before we summarize our findings in Section~\ref{sec:summary}.

\section{\bf Model and details of the numerical simulation}
\label{sec:model}

Our model system consists of $N$ point-particles situated in a quadratic simulation box in the $x,y$-plane with side-length $L$. The particles are 
uniform in mass~$m$ and charge~$q$ and are subject to periodic boundary conditions to avoid surface effects. 
The particles propagate according to the coupled equations of motion, 
\begin{equation}
m\ddot{\vec r}_i = \vec F_i + q \, \dot{\vec r}_i \times \vec B + \vec S_i\label{eq:eom},
\end{equation}
where the force $\vec F_i$ follows from the Yukawa potential created by all other particles, 
\begin{equation}
   \vec F_i = -\frac{q^2}{4\pi\varepsilon_0} \mathop{\sum\nolimits'}_{j=1}^N\left ( \nabla \frac{e^{-r/\lambda_D}}{r} \right )\Bigg \vert_{\vec r=\vec r_i - \vec r_j}. 
\end{equation}
Here, $\lambda_D$ denotes the Debye screening length and the primed sum indicates the omission of the term $j=i$.  
For $\lambda_D\rightarrow\infty$, we recover the well-known one-component plasma (OCP). 
The magnetic field is oriented perpendicular to the plane of the particles, $\vec B = B \vec e_z$, and 
the Langevin term $\vec S_i$ in Eq.~\eqref{eq:eom} is defined as
\begin{equation}
\vec S_i = - m_i \bar \nu \dot{\vec r}_i + \vec R_i\label{eq:stochastic}. 
\end{equation}
This Langevin term is only included in the simulations with friction $\bar \nu$, and 
$\vec R_i(t)$ is a Gaussian white noise with zero mean and the standard deviation
\begin{equation}
   \langle R_{\alpha,i}(t_0)R_{\beta,j}(t_0+t)\rangle=2k_BT\bar \nu\delta_{ij}\delta_{\alpha\beta}\,\delta(t),
\end{equation}
where $\alpha,\beta\in \{x,y\}$ and $T$ is the temperature.

Eq.~\eqref{eq:eom} is solved simultaneously for $N=4080$ particles using standard molecular dynamics simulation 
with integrators adopted to the influence of the magnetic field~\cite{Chin2008}. When friction is included in the simulations, we 
apply an additional Ornstein-Uhlenbeck process in momentum space~\cite{Forbert2000,Bussi2007}. For Coulomb systems, 
the appropriate Ewald summation techniques to calculate the forces are employed~\cite{Ewald1921,Parry1975}.
Starting from a random configuration of particles, 
the system is brought into equilibrium by a repeated rescaling of the particles' momenta according to the desired temperature $T$. 
After equilibrium conditions are realized, the system is advanced only according to Eq.~\eqref{eq:eom}.

In the following, lengths are given in units of the Wigner-Seitz radius $a=\left[n \pi \right]^{-1/2}$, where $n$ 
is the areal number density of the particles. Time is given in multiples of the inverse of the nominal angular 
plasma frequency $\omega_p=\left[ n q^2 /\left( 2\varepsilon_0 m a\right) \right]^{1/2}$. 
The system is characterized by four dimensionless parameters: 
\begin{enumerate}
 \item  the temperature $T$ of the particles, given in terms of the system's 
nominal (i.e., Coulomb-) coupling parameter $\Gamma=q^2/(4\pi\varepsilon_0 ak_BT)$,
\item the inverse of the Debye 
screening length $\lambda_D$, normalized by the Wigner-Seitz radius,  $\kappa = a/\lambda_D$,
\item the strength 
of the magnetic field, expressed as the ratio $\beta$ between the angular 
cyclotron frequency $\omega_c = qB/m$ and the nominal angular plasma frequency, 
$\beta=\omega_c/\omega_p$,
\item  the strength of friction, given by $\nu=\bar\nu/\omega_p$.

\end{enumerate}

\section{Collective Modes} 
\label{sec:collective}
The collective excitations of the strongly coupled, interacting particles can be analysed through the microscopic excitation spectra. 
The dispersion, i.e., the relation between the wavelength and the wave frequency of the collective 
excitations, can be obtained from the analysis of the autocorrelation (ACF) of the density and current fluctuations. 
For the Fourier components of the density fluctuations, we have~\cite{Hansen1975}
\begin{eqnarray}
\rho(\vec k, t) &=& \sum_{j=1}^N e^{i\vec k \cdot \vec r_j(t)}.  
\end{eqnarray}
The dynamical structure factor $S(\vec k, \omega)$ follows as
(assuming $\vec k=k\vec e_x$ and dropping the vector notation for~$\vec k$):
\begin{eqnarray}
S(k, \omega) &=& \frac{1}{2\pi N} \lim_{T\rightarrow\infty} \frac{1}{T} \left \vert \mathcal F_t\{\rho(k, t)\}\right \vert^2 \label{eq:sf}, 
\end{eqnarray}
where $\mathcal F_t$ denotes the temporal Fourier transform.

The current operator is given by~\cite{Hansen1975,Boon1991}
\begin{equation}
 \vec j(k, t) = \sum_{j=1}^N \vec v_j(t) \exp[{i k x_j(t)}]. 
\end{equation}
Separating longitudinal [$\lambda(k, t)$] and transverse [$\tau(k, t)$] currents, one arrives at the
microscopic quantities~\cite{footnote1}
\begin{eqnarray}
\lambda(k,t) &=&  \sum_{j=1}^N v_{jx} \exp[{ikx_j}],\label{eq:lambda}\\
\tau(k,t) &=&  \sum_{j=1}^N v_{jy} \exp[{ikx_j}],\label{eq:tau}
\end{eqnarray}
from which one can then obtain the fluctuation spectra $L(k,\omega)$ and $T(k,\omega)$ analogously to Eq.~\eqref{eq:sf} (replacing 
$\rho(k,t)$ by $\lambda(k, t)$ and $\tau(k, t)$, respectively):
\begin{eqnarray}
L(k, \omega) &=& \frac{1}{2\pi N} \lim_{T\rightarrow\infty} \frac{1}{T} \left \vert \mathcal F_t\{\lambda(k, t)\}\right \vert^2, \label{eq:lambdaf} \\
T(k, \omega) &=& \frac{1}{2\pi N} \lim_{T\rightarrow\infty} \frac{1}{T} \left \vert \mathcal F_t\{\tau(k, t)\}\right \vert^2 \label{eq:tauf}.
\end{eqnarray}

The dynamic structure factor $S(k, \omega)$ can be shown 
to be related to $L(k, \omega)$ as~\cite{Hansen2006}
\begin{equation}
 S(k,\omega) = \frac{k^2}{\omega^2} L(k, \omega). 
\end{equation}
$S(k, \omega)$, therefore, contains no additional information over $L(k,\omega)$, so we concentrate on $L(k, \omega)$ 
and $T(k, \omega)$ in the following~\cite{footnote2}. 

\begin{figure*}
 \includegraphics[scale=0.7]{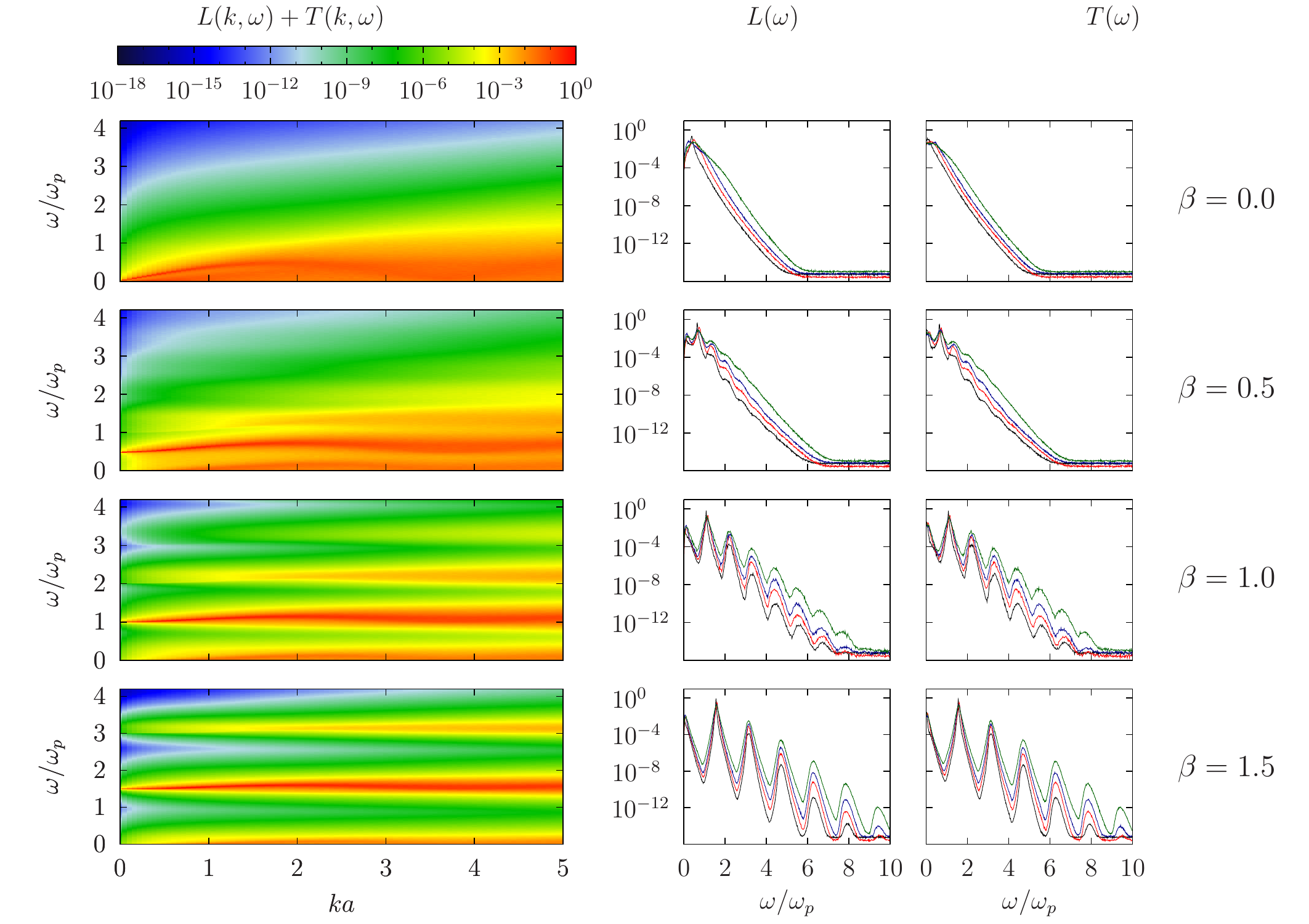}
\caption{(Color) $\kappa=2.0$, $\Gamma=200$, $\beta=0.0, 0.5, 1.0, 1.5$ (from top row to bottom row) Left: Collective excitation spectra, $L(k,\omega)+T(k,\omega)$, Right: $L(\omega)$ and $T(\omega)$
 at four values of $k$ ($ka=1, 2, 3, 5$, lowest to highest curve). 
}
\label{fig:var-b}
\end{figure*}

Our main interest lies in the mode spectra in the high-frequency range. We numerically evaluate the microscopic fluctuations 
\eqref{eq:lambda} and \eqref{eq:tau} 
at multiples of the minimum wavevector $\vec k_{\textrm{min}} = ({2\pi}/{L}) \vec e_x$ as dictated by our use of periodic 
boundary conditions. The data is subsequently Fourier analysed and the periodogram estimate of the power spectrum density 
is computed to obtain $L(k, \omega)$ and $T(k, \omega)$. Our simulations are typically thermalized for $20.000$ plasma periods 
and data are collected during $400.000$ or more plasma periods. 
The spectra are computed for different values and combinations of the coupling $\Gamma$, screening $\kappa$, magnetic 
field $\beta$ and friction $\nu$.

\section{Results}
\label{sec:results}

In this section, we study the generation of the higher harmonics in detail. Using more accurate 
data, we confirm the theoretical formula of the mode spacing reported in Ref.~\cite{Bonitz2010} in section~\ref{ssec:beta} and then discuss the effect 
of the coupling, the structural properties and the thermal energy on the higher harmonics in frictionless systems in section~\ref{ssec:temp}.
In section~\ref{ssec:dissipation}, we include dissipation and stochastic noise in our simulations to gauge the 
prospect of experimental verification.

\begin{figure}[t]
 \includegraphics[scale=0.6]{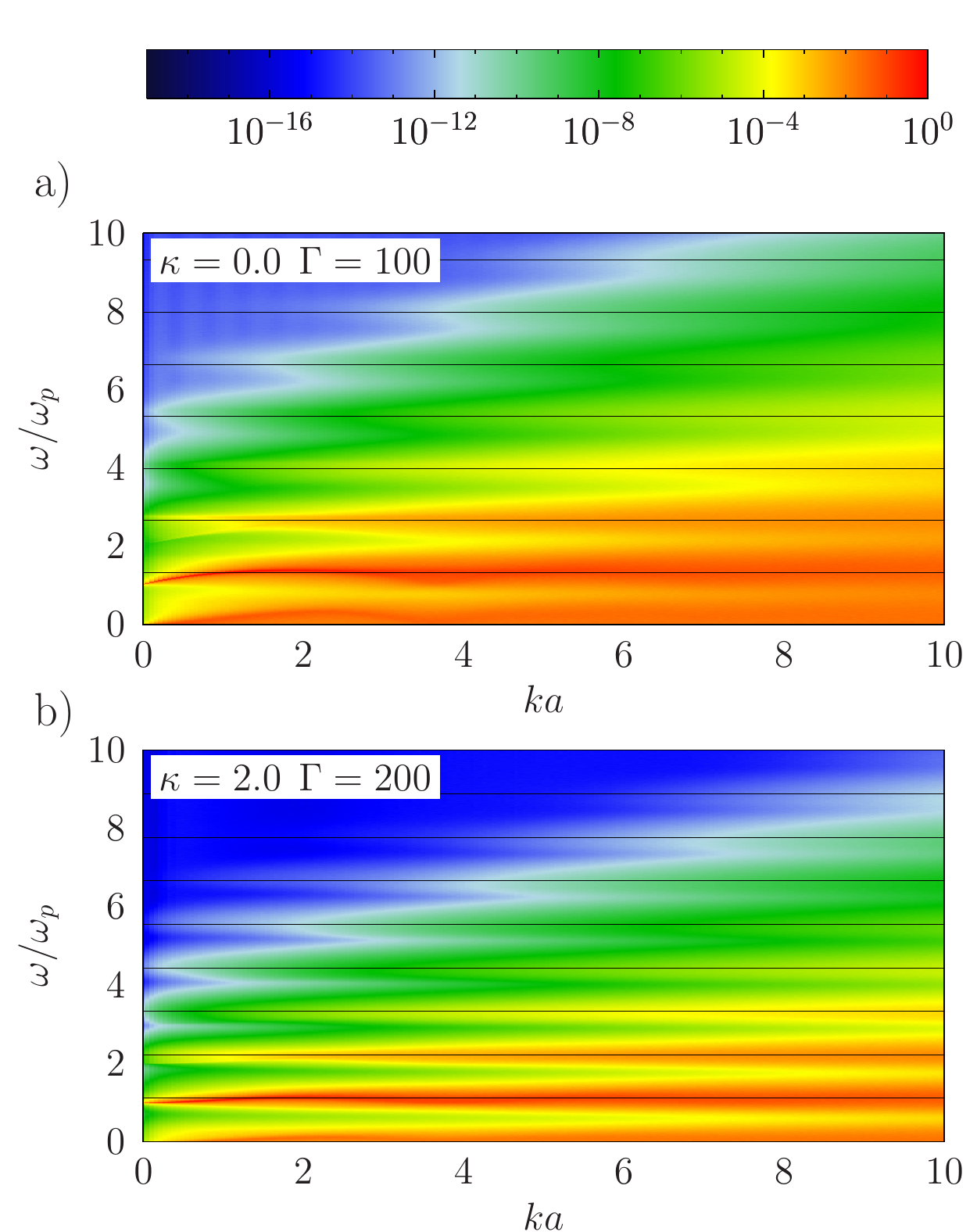}
\caption{(Color) Combined mode spectrum $L(k, \omega)+T(k, \omega)$ for a) $\kappa=0.0$, $\Gamma=100$ and b) $\kappa=2.0$, $\Gamma=200$, and $\beta=1.0$. Black 
lines are the theoretical prediction for higher harmonics according to Eq.~\eqref{eq:dbm}.}
\label{fig:g200b1}
\end{figure}

\begin{figure}
 \includegraphics[scale=0.6]{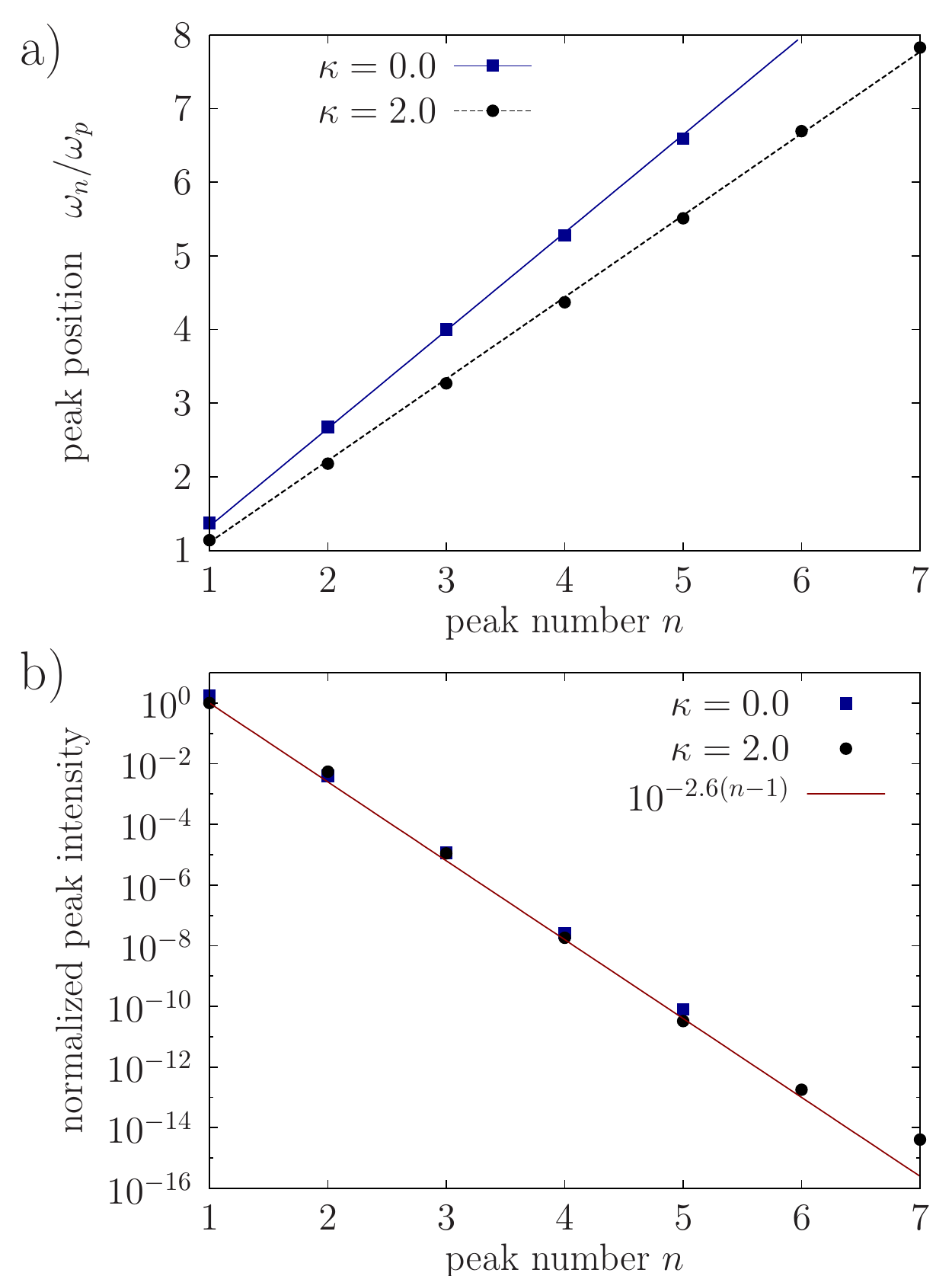}
\caption{$\kappa=2.0, \Gamma=200$, and $\kappa=0.0$, $\Gamma=100$ at  $\beta=1.0$, $ka=2.0$ 
\textit{a)}~The peak position of the $n$th mode. The straight lines are the theoretical predictions~\eqref{eq:dbm} with $\omega_E/\omega_0=0.62$ ($\kappa=0.0$) and $\omega_E/\omega_0=0.34$ ($\kappa=2.0$).
\textit{b)}~The peak intensity normalized to the intensity of the magnetoplasmon ($n=1$). An exponential decay is shown by the straight line. 
} 
\label{fig:g200b1_m}
\end{figure}

\begin{figure*}
 \includegraphics[scale=0.7]{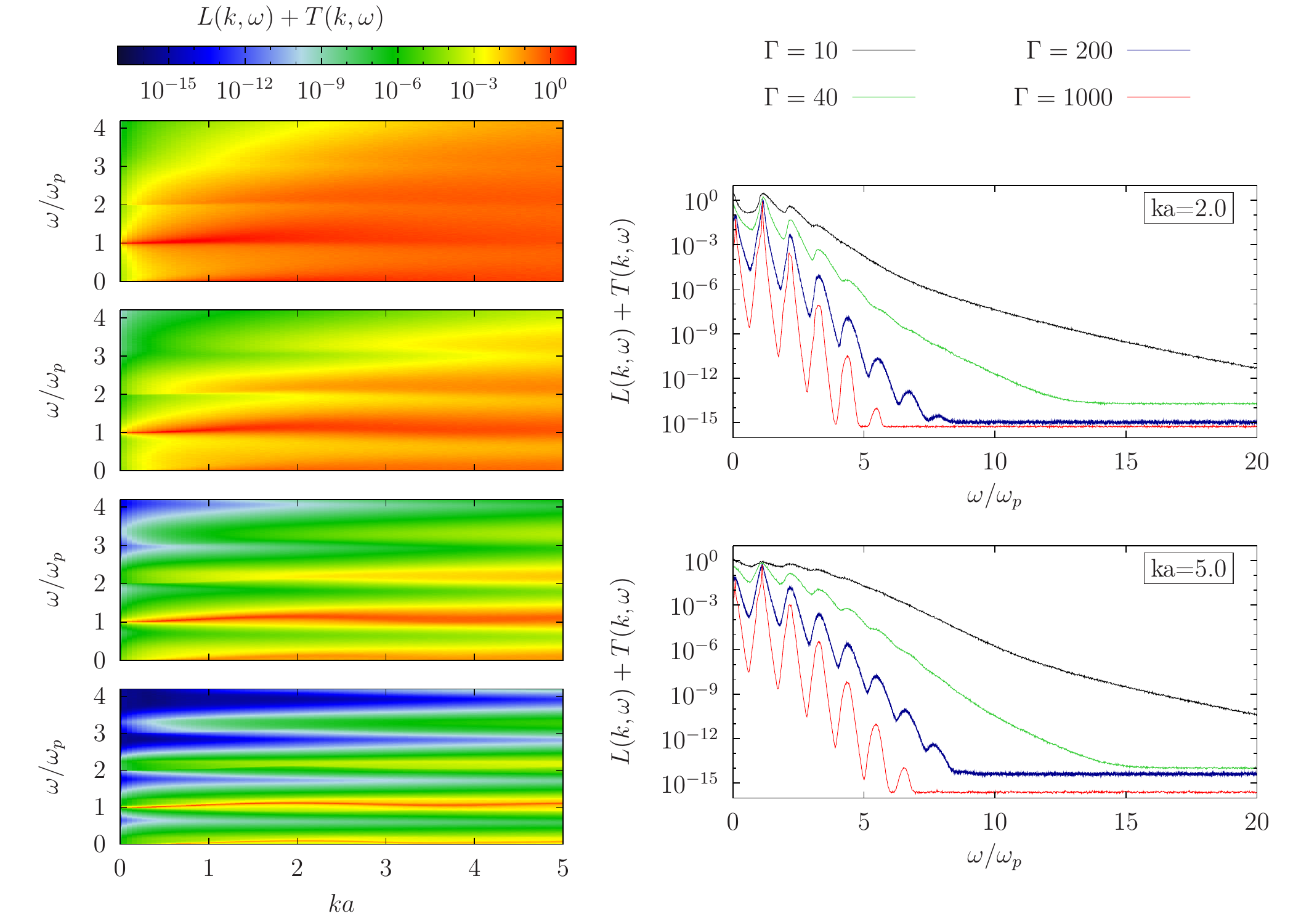}
\caption{(Color) $\beta=0.7$, $\kappa=2.0$, $\Gamma=10, 40, 200, 1000$,(from top to bottom) Left: Collective excitation spectra, $L(k,\omega)+T(k,\omega)$, Right: $L(k,\omega)+T(k,\omega)$ at two fixed values of $ka=2.0;5.0$ and values of $\Gamma$ as before. 
}
\label{fig:var-g_alt}
\end{figure*}

\subsection{Dependence of the spectrum on $\beta$}
\label{ssec:beta}
To obtain a first overview of the collective excitations, we depict in Fig.~\ref{fig:var-b} several results for the longitudinal 
and transverse wave spectra at different strengths of the magnetic field for a Yukawa system with $\kappa=2$ and $\Gamma=200$. 
The density plots in Fig.~\ref{fig:var-b} show the sum of transverse and longitudinal excitations. For the unmagnetized 
case (top row), the mode-spectrum is well-known and accurately described, e.g., within the QLCA~\cite{Golden2000}. 
The two branches associated with strongly coupled, unmagnetized Yukawa and Coulomb liquids (transverse and longitudinal branch) are not 
discernible on the scales of Fig.~\ref{fig:var-b}. At higher frequencies, the mode spectra appear structureless (cf. right columns of Fig.~\ref{fig:var-b}). 

In the the magnetized case (lower three rows of Fig.~\ref{fig:var-b}), two features are immediately noticeable: i) The two modes from 
the unmagnetized case are now replaced by the upper- and lower-hybrid modes (UH and LH) with the $k\rightarrow 0$ limits 
$\omega_0^\textrm{LH}=0$ and $\omega_0^\textrm{UH}=\omega_c$~\cite{Kalman2000}. 
ii) The qualitatively new feature reported in Ref.~\cite{Bonitz2010} is the emergence of multiple new branches 
at higher frequencies. These are higher harmonics reminiscent of the Bernstein modes which have been observed half a century ago in strongly magnetized {\em ideal plasmas} \cite{Bernstein1958}. Typical for these modes is that their frequencies are equally spaced, appearing at multiples of the cyclotron frequencies $\omega_c$ and that they are undamped and exist in the entire wave number range. 

The present modes show a number of fundamental differences which are caused by correlation effects: all modes are damped, and they appear only beyond a critical wave number which increases with the mode number. The most important difference is the mode frequency. Instead of being spaced by the cyclotron frequency $\omega_c$, here the modes appear at multiples of the magnetoplasmon frequency. 
The frequency of these ``dressed Bernstein'' modes, $\omega_n$, $n=2, 3, \dots$, were found to be well described by the relation~\cite{Bonitz2010}
\begin{equation}
\omega^2_n(k) \approx n \omega^2_{1, \infty},\qquad 
\omega^2_{1,\infty}(\beta,\kappa) = \omega^2_c(\beta) + 2\omega^2_E(\kappa).\label{eq:dbm}
\end{equation}
Here, the dominant single-particle oscillation frequency is denoted by $\omega_E$, the Einstein frequency, 
which strongly varies with $\kappa$ but also weakly depends on~$\Gamma$~\cite{Hartmann2005}. Some values 
of $\omega_E(\Gamma,\kappa)$ are collected in Tab.~\ref{tab:omega_E}.

\begin{table}[bp]
 \centering
\renewcommand{\arraystretch}{1.1}
\begin{tabular}{ccc}
\hline\hline
\phantom{xxx}$\kappa\phantom{xxx}$ &\phantom{xxx}$\Gamma$ \phantom{xxx} & \phantom{xxx}$\omega_E/\omega_p$\phantom{xxx} \\\hline
$0$ & 100 & 0.62  \\ 
$1$ & 150 & 0.52  \\ 
$2$ & 40 & 0.40  \\ 
$2$ & 100 & 0.36  \\ 
$2$ & 200 & 0.34  \\ 
$3$ & 600 & 0.20  \\ 

\hline
\end{tabular}
\caption{\label{tab:omega_E} The Einstein frequencies $\omega_E$ as calculated from QLCA~\cite{Donko2008} for various 
combinations of $\Gamma$ and $\kappa$. } 
\end{table}

In Ref.~\cite{Bonitz2010}, the data allowed to detect higher harmonics of the magnetoplasmon 
up to the third order[($n=4$ in Eq.~\eqref{eq:dbm}]. The present data show higher harmonics up to $n=7$ (cf. Fig.~\ref{fig:var-b}, third 
row). This provides us with the opportunity to verify the validity of Eq.~\eqref{eq:dbm} to a much higher precision. 
To this end, we depict in Fig.~\ref{fig:g200b1} the combined mode spectra for $\beta=1.0$ in two different systems. 
The solid lines indicate the theoretical prediction of Eq.~\eqref{eq:dbm}. 
A very good agreement between the theory and the simulations is evident. In addition, Fig.~\ref{fig:g200b1}(a) illustrates 
that the generation of the dressed Bernstein modes also occurs in a Coulomb system in which the particle interaction is long-ranged. 
This feature substantially expands the scope of the higher harmonics generation to other fields in plasma physics, including, e.g., 
ionic plasmas in traps~\cite{Dubin1999} and electron-hole plasmas in semiconductors. 

We now  quantify the higher harmonics in more detail. Fig.~\ref{fig:g200b1_m} shows the frequency and oscillator 
strengths of the higher harmonics and the magnetoplasmon at $ka=2.0$ and fixed $\Gamma$ and~$\beta$. The agreement 
between relation~\eqref{eq:dbm} and the peak position from the simulations is again excellent (Fig.~\ref{fig:g200b1_m}a), 
both for the OCP and a typical Yukawa system with $\kappa=2$. 
Note that no free parameter enters the theoretical prediction and that the differences in the theoretical predictions 
are only mediated by $\omega_E$. 

A quantity of central interest is the relative intensity of the generated higher harmonics. These values provide one
with a first estimate of the required experimental sensitivity to observe the described effects. 
Figure~\ref{fig:g200b1_m}b shows the relative intensities at $ka=2.0$ for the Yukawa and Coulomb spectra. 
Evidently, the data points are well described by a decay of the form $10^{-2.6(n-1)}$. 
The first of the higher harmonics (i.e., $n=2$)  appears with 
an intensity of about one hundredth of the magnetoplasmon. Such intensities are observable against the background noise 
in typical dusty plasma experiments~\cite{Nunomura2005}. 

\subsection{Dependence of the spectrum on the interaction, temperature, and structural order}
\label{ssec:temp}
The frictionless Yukawa system is (at fixed magnetization $\beta$) characterized by two parameters: The inverse temperature 
(Coulomb coupling parameter) $\Gamma$ and the interaction range (inverse Debye length) $\kappa$. The influence 
of the temperature is investigated by changing, at a fixed value of $\kappa$, the coupling 
parameter~$\Gamma$. 
%

In Fig.~\ref{fig:var-g_alt}, 
different situations are depicted for a system with $\kappa=2$, ranging from intermediate coupling ($\Gamma=10;40$) 
to strong ($\Gamma=200$) and very strong coupling ($\Gamma=1000$). Note that for $\Gamma=1000$, the system is already 
far in the microcrystalline regime, $\Gamma_\text{rel}=\Gamma/\Gamma_\text{melting}\approx~2.4$. 
The decreased thermal background makes the fundamental and higher harmonics stand out very clearly at strong coupling, and the 
signal-to-noise ratio is higher in these cases. For example, harmonics with $n>3$ are far easier to distinguish at $\Gamma=1000$ than at $\Gamma=40$. The intensity ratio 
$M=I_n/I_0$ between the intensity of the $n$th harmonic and the fundamental is, however, clearly decreased for stronger 
coupling~\cite{footnote3}. 
This indicates that a certain degree of disorder is required for an effective 
generation of higher harmonics, which in turn might depend on anharmonic effects that are less important in highly ordered 
systems. 

\begin{figure}
 \includegraphics[scale=0.6]{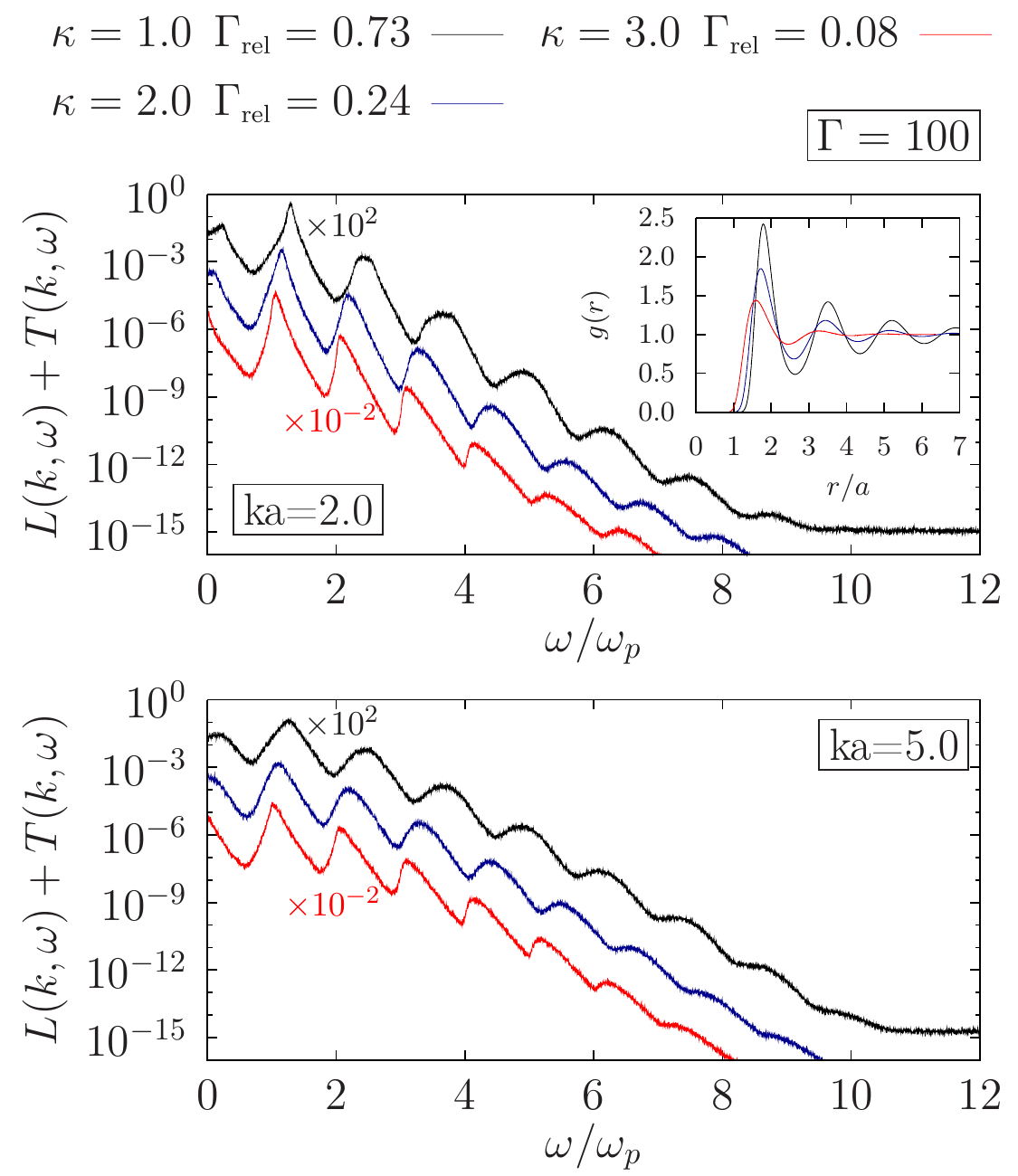}
\caption{(Color) Combined mode spectra at $ka=2.0; 5.0$ for identical temperature ($\Gamma=100$) and $\kappa=1.0; 2.0; 3.0$
(highest to lowest curve), $\beta=1.0$. The curves are shifted vertically for clarity. 
The inset shows the RPDF $g(r)$. }
\label{fig:g100}
\end{figure}

Having established the influence of the temperature (inverse coupling), we now determine the influence of the interaction range. 
To this end, in Fig.~\ref{fig:g100}, we show the combined mode spectra at a fixed $\Gamma=100$ for different values of $\kappa$ and two 
values of $ka$. 
The curves appear very similar, with comparable signal-to-noise-ratios and similar
relative intensities of the higher harmonics. This is despite the strong differences in structural order among the systems 
reflected in the different radial pair distribution functions (RPDF, inset of Fig~\ref{fig:g100}) defined as
\begin{equation}
g(\vert \vec r\vert)=\frac{L^2}{N^2}\left\langle \mathop{\sum\nolimits'}_{i,j}\delta(\vert \vec r - \vec r_{ij}\vert)\right\rangle.
\end{equation}
The sole effect of an increase in $\kappa$ on the higher harmonics is a systematic shift toward smaller 
frequencies. This result is not unexpected and indeed predicted by Eq.~\eqref{eq:dbm}: With increasing $\kappa$, the 
particle oscillations decrease in frequency as the interparticle potential becomes more lenient. The Einstein frequency, therefore, 
diminishes with increasing $\kappa$ (cf. Table~\ref{tab:omega_E}) and, consequently, so do the frequencies at which 
the higher harmonics appear. 

So far we have investigated the influence of $\Gamma$ and $\kappa$ on the higher harmonics. A change of either of these 
parameters also affects the structural properties of the system. By changing $\Gamma$ and $\kappa$ simultaneously in such a way 
that the RPDF remains practically unchanged~\cite{Vaulina2002a, Hartmann2005, Vaulina2006, Ott2010a}, we can compare systems with identical structural properties.
We achieve this by fixing the ratio of the temperature to the point of (micro-)crystallization, 
$\Gamma/\Gamma_\text{melting}\equiv \Gamma_\text{rel}$~\cite{Liu2007,Ott2009a}. Systems corresponding to the same $\Gamma_\text{rel}$ 
but different $\Gamma$ have thus the same structural order but different thermal energy.

\begin{figure}
 \includegraphics[scale=0.6]{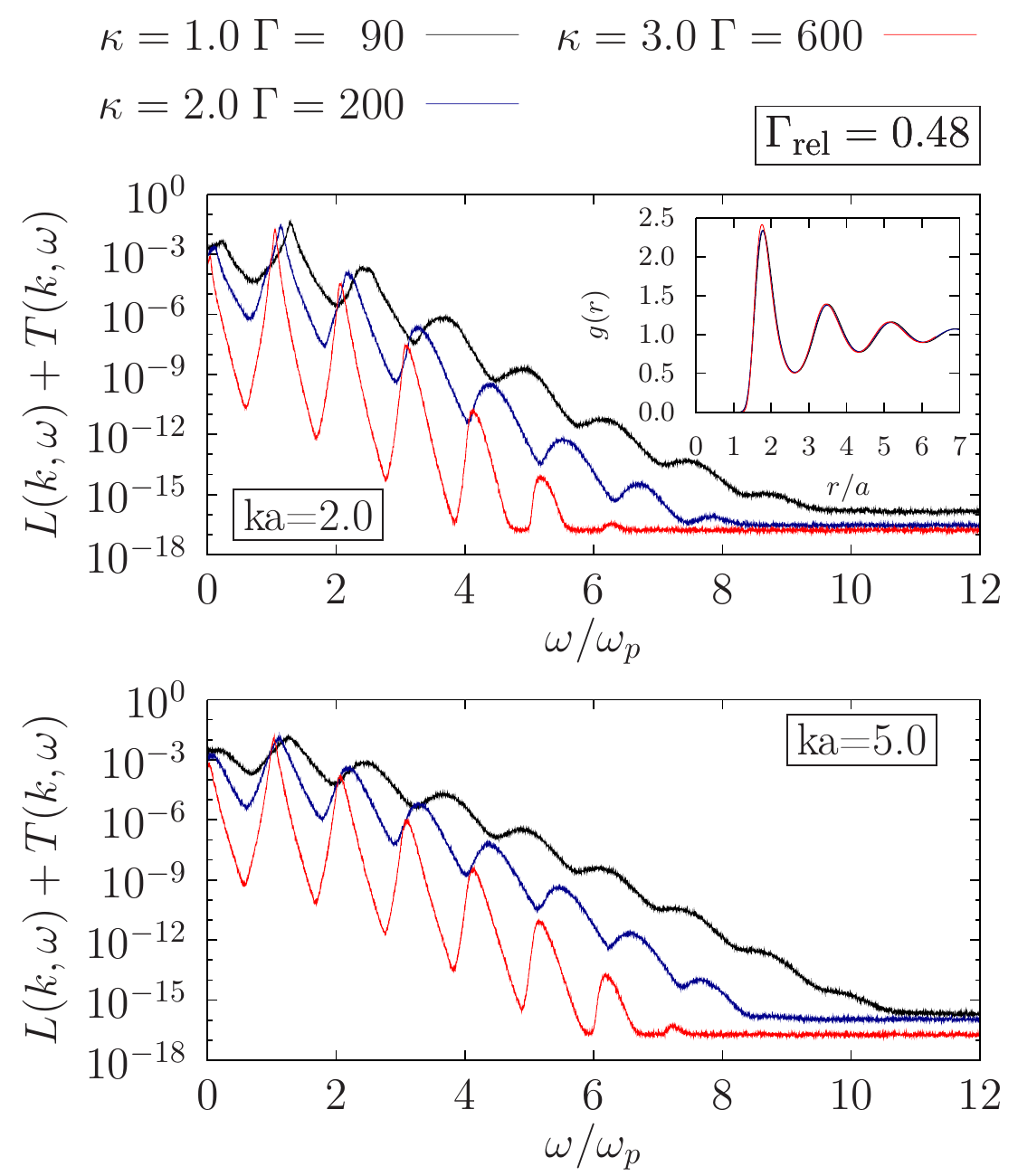}
\caption{(Color) Combined mode spectra at $ka=2.0; 5.0$ for identical structural properties 
corresponding to $\Gamma_\text{rel}=0.48$ but different $\kappa=1.0; 2.0; 3.0$ 
(highest to lowest curve), $\beta=1.0$. 
The inset shows the RPDF $g(r)$. }
\label{fig:geff048}
\end{figure}

The results are presented in Fig.~\ref{fig:geff048}, where a fixed value of $\Gamma_\textrm{rel}$ is maintained for systems with 
different values of~$\kappa$. A combination of the two previously described effects is observed: The higher harmonics 
shift toward lower frequency with increasing~$\kappa$, and decay more slowly albeit with a decreased signal-to-noise-ratio with increasing 
temperature (inverse coupling). 

In conclusion, we find that the generation 
of higher harmonics is nearly independent of the intricate details of the structure and the range of the 
interaction between the particles (except for a systematic frequency shift) but, instead, 
is dominated by the thermal energy available to the system.

\subsection{Influence of dissipation on the higher harmonics. Prospects for experimental observation}
\label{ssec:dissipation}
\begin{figure}
 \includegraphics[scale=0.6]{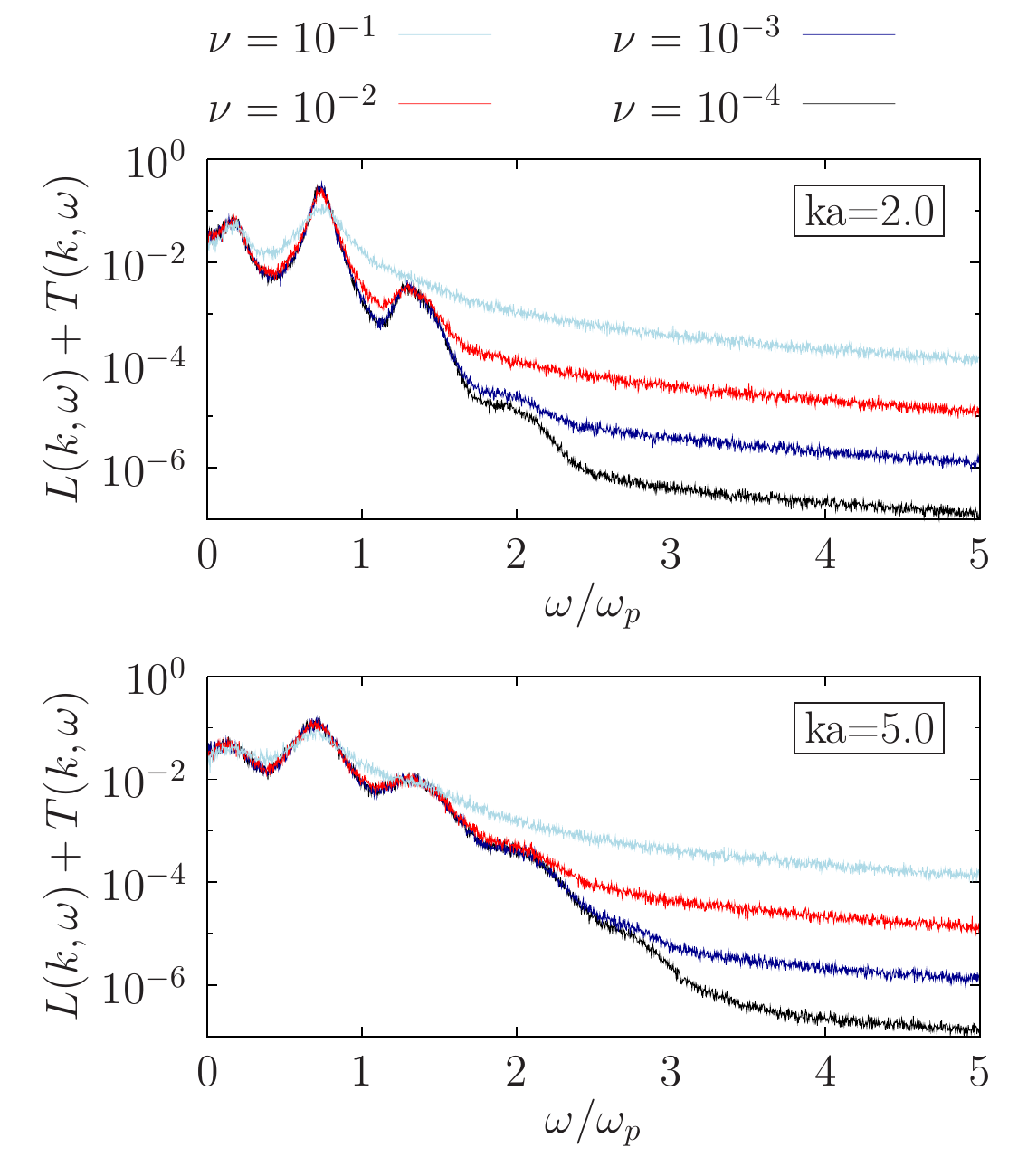}
\caption{$\Gamma=200$, $\kappa=2.0$, $\beta=0.5$: Combined mode spectra for different values of the friction strength, 
from $\nu=10^{-4}$ (top curve) to $\nu=10^{-1}$ (lowest curve). 
} 
\label{fig:g200b05_v}
\end{figure}

Besides noise resulting from the measurement itself, many physical systems are also influenced by 
fluctuations due to the coupling to their surroundings. In many cases, these fluctuations  and dissipation effects are well described by 
a constant friction and stochastic noise, an approach which has been used extensively in the description of, e.g., 
dusty plasmas. We now introduce dissipation and stochastic noise in our MD simulations to estimate the effect of friction-induced noise on 
the fluctuation spectra. In other words, we now include the term~\eqref{eq:stochastic} in Eq.~\eqref{eq:eom}. 

In Fig.~\ref{fig:g200b05_v}, we depict the longitudinal fluctuation spectrum $L(k,\omega)$ for fixed values of $k$, $\beta$ and $\Gamma$, 
while varying the magnitude of the friction $\nu$. As expected, the inclusion of friction has a two-fold effect: 
a) The overall noise level in the spectrum is increased and begins to overlap the higher harmonics and b) the peaks in the spectra become increasingly broader with higher friction. 

The noise resulting from the friction clearly dominates the intrinsic noise for the data depicted in Fig.~\ref{fig:g200b05_v}. 
The high-$\omega$ limit of the spectra increases linearly with the friction coefficient.
From the data presented in Fig.~\ref{fig:g200b05_v}, we conclude 
that at a moderately coupled liquid state, $\Gamma=200$ and $\kappa=2$, the detection of the second harmonic is possible 
at $\beta=0.5$ and a friction of $\nu \sim 0.01$ (note that $\nu$ is given here in units of the nominal plasma frequency). 

We now give estimates for the plasma parameters required to observe the generation of higher harmonics under experimental 
conditions. The main obstacle in attaining our simulation conditions in experiments are the required high magnetic fields. 
Combining the formulas for the cyclotron frequency $\omega_c$ and the plasma frequency $\omega_p$ and solving 
for the magnetic field, we obtain (in SI units): 
\begin{equation*}
B[\textrm T] \approx 2.7\cdot 10^5\beta  \left( \frac{R}{a}\right)^\frac{3}{2}\sqrt{\rho\left[\frac{\textrm{kg}}{\textrm m^3}\right]},
\end{equation*}
where $\rho$ is the mass density. 
Typical 2D dusty plasma parameters are $R=1\dots 10 \mu\textrm{m}$, $\rho\approx 1\textrm{g}/\textrm{cm}^{3}$ and $a=0.1\dots 1 \textrm{mm}$, giving 
rise (for $\beta\approx 0.074$) to a required magnetic field strength of about $20\textrm T$, which is at the edge of current 
experimental possibilities. That number can be reduced by using 
lighter particles or compressing the system to decrease the inter-particle spacing (an increase of the average density 
by a factor of, e.g., $2$ lowers the magnetic field requirements by 40\%). Also note that the required magnetic field 
is independent of particle charge, which, however, enters into the coupling parameter $\Gamma$. 

The dissipation in 2D dusty plasma experiments can be quite low, a typical value is $\nu \approx 0.02$~\cite{Liu2008a}. Thus, 
according to our simulations, friction  at this rate should not prevent the generation of the higher harmonics. It is, however, not 
uncommon to experience friction rates in excess of $\nu = 0.1$, at which the observation of the higher harmonics 
is not possible. 
In conclusion, the observation of higher harmonics appears as a possible experimental venue but poses several 
challenging restraints on the experimental setup.

\section{Summary}
\label{sec:summary}
In this work, we have compared the previously derived theoretical description of the mode spacing of the higher harmonics with extended MD simulations and found excellent agreement over a wide range of coupling strengths and interaction range. We have also, for the first time, demonstrated that the generation of higher harmonics is present in strongly correlated systems with Coulomb interaction as well, which is especially interesting in the light of additional possibilities for experimental verification. 
The relative intensities of the higher harmonics have also been investigated and were found to decay, to a very good approximation, exponentially with the order of the harmonics. 
Finally, we have included dissipation in our simulations and estimated the maximum acceptable level of such friction for experiments. Combining estimates for the strength of the magnetic field and the maximum friction levels, we conclude that the experimental observation of higher harmonics is possible but challenging. 

Finally, we expect that our results are also of direct relevance for strongly correlated quantum plasmas. In the case of magnetized ideal quantum plasmas the ``traditional'' Bernstein modes have been predicted theoretically long ago by Horing et al. \cite{horing_1976} and were observed experimentally in electron-hole plasmas, e.g. \cite{klitzing_2002}. In recent years strongly correlated quantum Coulomb systems, including liquid states and Wigner crystals of electrons \cite{filinov_prl01}, holes \cite{bonitz_prl05} and excitons, e.g. \cite{ludwig_pss06}, moved into the focus of research. It is expected that these systems, if placed into a strong magnetic field, should exhibit a similar spectrum of ``dressed Bernstein modes'' as their classical counterpart studied in the present paper.

\begin{acknowledgements}
This work is supported by the Deutsche Forschungsgemeinschaft via SFB-TR 24 (projects A5 and A7), Hungarian Grants OTKA-T-77653, 
OTKA-PD-75113, the J\'anos Bolyai Research Foundation of the Hungarian
Academy of Sciences, and by the North-German Supercomputing Alliance (HLRN) via grant shp0006. One of us (TO) gratefully acknowledges helpful discussions with SA Chin. 
\end{acknowledgements}


\begin{thebibliography}{31}
\bibitem{footnote1} Note that some authors insert a factor $k$ in front of the sums in Eqs.~\eqref{eq:lambda} and~\eqref{eq:tau}.
\bibitem{footnote2} Note that the displayed spectra are scaled for maximum visibility and are given in arbitrary units throughout this work.
\end{thebibliography}
\end{document}